\documentclass[12pt]{article}
\usepackage{amsmath}
\usepackage{amssymb}
\usepackage{amsfonts}
\usepackage [latin1]{inputenc}

\numberwithin{equation}{section}

\title{
DYNAMICAL AND INVARIANCE ALGEBRAS OF THE $d$-DIMENSIONAL DUNKL-COULOMB PROBLEM}
\author{C. QUESNE\\ 
{\small D\'epartement de Physique,  Universit\'e Libre de Bruxelles,} \\ 
{\small Campus de la Plaine CP229, Boulevard~du Triomphe, B-1050 Brussels, Belgium}\\
{\small (e-mail: Christiane.Quesne@ulb.be})}
\date{ }
\begin{document}
\baselineskip=22pt plus 1pt minus 1pt
\maketitle

\begin{abstract} 
It is shown that the rich algebraic structure of the standard $d$-dimensional Coulomb problem can be extended to its Dunkl counterpart. Replacing standard derivatives by Dunkl ones in the so($d+1$,2) dynamical algebra generators of the former gives rise to a deformed algebra with similar commutation relations, except that the metric tensor becomes dependent on the reflection operators and that there are some additional commutation or anticommutation relations involving the latter. It is then shown that from some of the dynamical algebra generators it is straightforward to derive the integrals of motion of the Dunkl-Coulomb problem in Sturm representation. Finally, from the latter, the components of a deformed Laplace-Runge-Lenz vector are built. Together with the Dunkl angular momentum components, such operators insure the superintegrability of the Dunkl-Coulomb problem in Schr\"odinger representation. 
\end{abstract}

\newpage
\section{Introduction}

In 1950, Wigner introduced a deformed version of the Heisenberg algebra in quantum mechanics \cite{wigner} and soon after that Yang proposed the reflection operator to discuss the harmonic oscillator problem along these lines \cite{yang}. Dunkl independently considered sets of differential-difference operators associated with finite reflection groups \cite{dunkl89}. These operators are now referred to as Dunkl operators and are very useful for studying polynomials in several variables with discrete symmetry \cite{dunkl14}.\par
%
%
In quantum mechanics, Dunkl operators have been getting a lot of applications too, so that they have given rise to what is now often called Wigner-Dunkl quantum mechanics. One may quote, for instance, their use for bosonizing supersymmetric quantum mechanics \cite{plyu} or generalizations thereof \cite{cq21}, for describing an exchange operator formalism in Calogero-Moser-Sutherland models and their generalizations \cite{brink, poly, cq95}, and for proving the superintegrability of some models \cite{cq10}.\par
%
%
More recently, some specific exactly-solvable quantum systems wherein the ordinary derivatives are replaced by Dunkl ones have been worked out. Much interest has been devoted, for instance, to the Dunkl oscillators in one, two, and three dimensions \cite{genest13a, genest14a, genest13b, genest14b}, as well as the Dunkl-Coulomb problems in two and three dimensions \cite{genest15, gha19, gha20}. One may also quote the introduction in Wigner-Dunkl quantum mechanics of some rationally-extended potentials, whose wavefunctions can be written in terms of exceptional orthogonal polynomials \cite{cq23a, cq23b}, and the construction of quasi-exactly solvable potentials \cite{cq24}.\par
%
%
The purpose of the present paper is to extend the rich algebraic structure of the standard Coulomb problem (for a review see, e.g.\ \cite{wybourne}) to its Dunkl counterpart. We plan to start from a deformed version of the well-known so($d+1$,2) dynamical algebra of the $d$-dimensional standard Coulomb problem (for the three-dimensional case see, e.g., \cite{barut}) and to show that such an approach leads to a simple derivation of the invariance algebra of the $d$-dimensional Dunkl-Coulomb problem in Sturm representation. From the latter, we will then determine the invariance algebra of such a problem in Schr\"odinger representation. In this way, we will generalize the standard Laplace-Runge-Lenz (LRL) vector, whose introduction in quantum mechanics dates back to Pauli \cite{pauli} and which explains the superintegrability of the Coulomb problem.
\par
%
%
The paper is organized as follows. In Section~2, we introduce the $d$-dimensional Dunkl-Coulomb problem. Its dynamical algebra is defined in Section~3. In Section~4, we determine the invariance algebra of such a problem in Sturm representation. In Section~5, this invariance algebra is transformed into that in Schr\"odinger representation. Finally, Section~6 contains the conclusion. \par
%
%
\section{\boldmath $d$-dimensional Dunkl-Coulomb problem}

The $d$-dimensional Dunkl-Coulomb model is defined by the Hamiltonian
\begin{equation}
  H = - \frac{1}{2} \mathbf{D}^2 - \frac{\alpha}{r},  \label{eq:H}
\end{equation}
where $r = \left(\sum_{i=1}^d x_i^2\right)^{1/2}$, $\alpha>0$, and
\begin{equation}
  \mathbf{D}^2 = \sum_{i=1}^d D_i^2
\end{equation}
is the $d$-dimensional Dunkl-Laplacian operator, defined in terms of the Dunkl derivatives
\begin{equation}
  D_i = \partial_i + \frac{\mu_i}{x_i}(1 - R_i), \qquad i=1, 2, \ldots, d, \qquad \mu_i>0.
\end{equation}
Here, $R_i$ is the reflection operator defined by
\begin{equation}
  R_i f(x_i) = f(-x_i)
\end{equation}
and $\partial_i f = \partial f/\partial x_i$. For future purposes, we note that
\begin{align}
  &[D_i, x_j] = \delta_{i,j} (1+2\mu_i R_i), \nonumber \\
  &\{R_i,x_i\} = \{R_i, D_i\} = 0, \label{eq:Dunkl-com} \\
  &[R_i, x_j] = [R_i, D_j] = [R_i, R_j] = 0 \qquad \text{if $i\ne j$}. \nonumber  
\end{align}
\par
%
%
It is also worth observing that the change of standard derivatives into Dunkl ones is accompanied by a change of scalar product, which is now defined by
\begin{equation}
  <g | f> = \int_{-\infty}^{\infty} \int_{-\infty}^{\infty} \cdots \int_{-\infty}^{\infty} g^*(\mathbf{x})
  f(\mathbf{x}) |x_1|^{2\mu_1} |x_2|^{2\mu_2} \cdots |x_d|^{2\mu_d} dx_1 dx_2 \cdots dx_d.   
  \label{eq:sp}
\end{equation}
\par
%
%
The Schr\"odinger equation corresponding to (\ref{eq:H}) reads
\begin{equation}
  H \Psi(\mathbf{x}) = \left(-\frac{1}{2} \mathbf{D}^2 - \frac{\alpha}{r}\right) \Psi(\mathbf{x}) =
  E \Psi(\mathbf{x}).
\end{equation}
Instead of the latter, one may consider the eigenvalue problem
\begin{equation}
  K \Psi(\mathbf{x}) = \alpha \Psi(\mathbf{x}). 
\end{equation}
for the operator $K$ defined by
\begin{equation}
  K = - \frac{r}{2} \mathbf{D}^2 - E r. \label{eq:K}
\end{equation}
This corresponds to the so-called Sturm representation for the Coulomb problem, wherein the energy $E$ is considered fixed, while $\alpha$ becomes the eigenvalue.\par
%
%
\section{Dynamical algebra of the Dunkl-Coulomb problem}

\setcounter{equation}{0}

To start with, let us introduce three operators generalizing some operators known for the three-dimensional standard Coulomb problem \cite{barut, dhoker}
\begin{equation}
  \Gamma_0 = \frac{1}{2} r(-\mathbf{D}^2+1), \quad \Gamma_{d+1} = \frac{1}{2} r(-\mathbf{D}^2-1),
  \quad T = - {\rm i} \biggl(\mathbf{r}\cdot\mathbf{D} + \frac{d-1}{2} + \sum_i \mu_i R_i\biggr).
  \label{eq:so(2,1)}
\end{equation}
It is straightforward to show that they satisfy the commutation relations
\begin{equation}
  [\Gamma_0, \Gamma_{d+1}] = {\rm i} T, \qquad [\Gamma_0, T] = - {\rm i} \Gamma_{d+1}, \qquad
  [\Gamma_{d+1}, T] = - {\rm i} \Gamma_0,  \label{eq:com-1}
\end{equation}
showing that they generate an so(2,1) algebra.
\par
%
%
Let us now introduce the components of the Dunkl angular momentum in $d$ dimensions,
\begin{equation}
  J_{ij} = - J_{ji} = - {\rm i} (x_i D_j - x_j D_i), \qquad i, j = 1, 2, \ldots, d.  \label{eq:J}
\end{equation}
Its square can be shown to be given by
\begin{equation}
  \mathbf{J}^2 = \frac{1}{2} \sum_{ij} J_{ij}^2 = - \mathbf{x}^2 \mathbf{D}^2 + (\mathbf{x} \cdot 
  \mathbf{D})^2 + \mathbf{x} \cdot \mathbf{D} \biggl(d-2+2\sum_i \mu_i R_i\biggr).
  \label{eq:J-2}
\end{equation}
On comparing the latter with the Casimir operator of the so(2,1) algebra generated by $\Gamma_0$, $\Gamma_{d+1}$, and $T$, namely
\begin{equation}
  Q^2 = \Gamma_0^2 - \Gamma_{d+1}^2 - T^2,
\end{equation}
we get the relation
\begin{equation}
  Q^2 = \mathbf{J}^2 + \biggl(\frac{d-3}{2} + \sum_i \mu_i R_i\biggr) \biggl(\frac{d-1}{2}
  + \sum_j \mu_j R_j\biggr). 
\end{equation}
\par
%
%
Let us also consider three sets of $d$ operators, defined by
\begin{equation}
\begin{split}
  A_i &= - \frac{1}{2} x_i \mathbf{D}^2 + D_i \biggl(\mathbf{x} \cdot \mathbf{D}+ \frac{d-3}{2} 
     + \sum_{j} \mu_j R_j\biggr) - \frac{1}{2} x_i, \\
  M_i &= - \frac{1}{2} x_i \mathbf{D}^2 + D_i \biggl(\mathbf{x} \cdot \mathbf{D}+ \frac{d-3}{2}
     + \sum_{j} \mu_j R_j\biggr) + \frac{1}{2} x_i, \\
  \Gamma_i &= - {\rm i} r D_i,
\end{split}. \label{eq:gen}
\end{equation}
with $i=1, 2, \ldots, d$. \par
%
%
The commutation relations of the so(2,1) generators (\ref{eq:so(2,1)}) with the four sets of operators defined in (\ref{eq:J}) and (\ref{eq:gen}) are easily obtained in the form
\begin{equation}
\begin{split}
  &[\Gamma_0, J_{ij}] = [\Gamma_{d+1}, J_{ij}] = [T, J_{ij}] = 0, \\
  &[\Gamma_0, A_i] = [\Gamma_{d+1}, M_i] = [T, \Gamma_i] = 0, \\
  &[\Gamma_{d+1}, A_i] = - [\Gamma_0, M_i] = {\rm i} \Gamma_i, \\
  &[T, A_i] = [\Gamma_0, \Gamma_i] = {\rm i} M_i, \\
  &[T, M_i] = [\Gamma_{d+1}, \Gamma_i] = {\rm i} A_i,
\end{split}. \label{eq:com-2}
\end{equation}
and we note that they coincide with those obtained for the corresponding operators in the standard Coulomb problem, i.e., for $\mu_i = 0$ \cite{barut, dhoker}.\par
%
%
On finally considering the commutation relations of operators (\ref{eq:J}) and (\ref{eq:gen}) among themselves, we obtain some relations that remain the same as for the standard Coulomb problem, namely
\begin{equation}
  [A_i, A_j] = - [M_i, M_j] = - [\Gamma_i, \Gamma_j] = {\rm i} J_{ij},  \label{eq:com-3}
\end{equation}
as well as some relations that are changed by the appearance of some reflection operator on the right-hand side, namely
\begin{equation}
\begin{split}
  &[J_{ij}, J_{kl}] = {\rm i} [\delta_{i,k} J_{jl} (1 + 2\mu_k R_k) + \delta_{i,l} J_{kj} (1 + 2\mu_l R_l) \\
  &\phantom{[J_{ij}, J_{kl}]}\quad + \delta_{j,k} J_{li} (1 + 2 \mu_k R_k) + \delta_{j,l} J_{ik} 
      (1+ 2 \mu_l R_l)],\\
  &[J_{ij}, A_k] = {\rm i} (\delta_{i,k} A_j - \delta_{j,k} A_i) (1 + 2\mu_k R_k), \\
  &[J_{ij}, M_k] = {\rm i} (\delta_{i,k} M_j - \delta_{j,k} M_i) (1 + 2\mu_k R_k), \\
  &[J_{ij}, \Gamma_k] = {\rm i} (\delta_{i,k} \Gamma_j - \delta_{j,k} \Gamma_i) (1 + 2\mu_k R_k), \\
  &[A_i, M_j] = {\rm i} \delta_{i,j} T (1 + 2\mu_i R_i), \\
  &[A_i, \Gamma_j] = {\rm i} \delta_{i,j} \Gamma_{d+1} (1 + 2\mu_i R_i), \\
  &[M_i, \Gamma_j] = {\rm i} \delta_{i,j} \Gamma_0 (1 + 2\mu_i R_i).
\end{split}. \label{eq:com-4}
\end{equation}
\par
%
%
In addition, we have some commutation or anticommutation relations involving reflections,
\begin{equation}
\begin{split}
  &[R_i, \Gamma_0] = [R_i, \Gamma_{d+1}] = [R_i, T] = 0, \\
  &[R_i, J_{jk}] = 0 \qquad \text{if $i\ne j,k$}, \\
  &\{R_i, J_{jk}\} = 0 \qquad \text{if $i=j$ or $i=k$}, \\  
  &[R_i, A_j] = [R_i, M_j] = [R_i, \Gamma_j] = 0 \qquad \text{if $i\ne j$}, \\
  &\{R_i, A_j\} = \{R_i, M_j\} = \{R_i, \Gamma_j\} = 0 \qquad \text{if $i=j$}.
\end{split}. \label{eq:com-R}
\end{equation}
\par
%
%
In the standard Coulomb problem, the $\frac{1}{2}(d+2)(d+3)$ operators $\Gamma_0$, $\Gamma_{d+1}$, $T$, $J_{ij} = - J_{ij}$ ($1 \le i < j \le d$), and $A_i$, $M_i$, $\Gamma_i$, ($i=1, 2, \ldots, d$) can be identified with the generators $L_{ab} = - L_{ba} = L_{ab}^{\dagger}$, $a, b = 1, 2, \ldots, d+3$, of an so($d+1$,2) algebra, whose commutation relations are given by
\begin{equation}
  [L_{ab}, L_{cd}| = {\rm i} (g_{ac} L_{bd} + g_{ad} L_{cb} + g_{bc} L_{da} + g_{bd} L_{ac}),
  \label{eq:so(d+1,2)}
\end{equation}
where $g_{ab} = {\rm diag}(1, 1, \ldots, 1,-1,-1)$. The identification is the following one:
\begin{equation}
\begin{split}
  &L_{ij} = J_{ij}, \qquad i,j= 1,2,\ldots,d, \\
  &L_{i,d+1} = A_i, \qquad L_{i,d+2}= M_i, \qquad L_{i,d+3} = \Gamma_i, \qquad i=1, 2, \ldots, d, \\
  &L_{d+1,d+2} = T, \qquad L_{d+1,d+3} = \Gamma_{d+1}, \qquad L_{d+2,d+3} = \Gamma_0.
\end{split}
\end{equation}
\par
%
%
With the same identification, it turns out from (\ref{eq:com-1}), (\ref{eq:com-2}), (\ref{eq:com-3}), and (\ref{eq:com-4}) that the operators (\ref{eq:so(2,1)}), (\ref{eq:J}), and (\ref{eq:gen}), obtained by changing standard derivatives into Dunkl ones, still satisfy an equation of type (\ref{eq:so(d+1,2)}) provided we interpret $g_{ab}$ as the operator-dependent diagonal matrix $g_{ab} = {\rm diag}(1+2\mu_1R_1, 1+2\mu_2R_2, \ldots, 1+2\mu_dR_d, 1, -1, -1)$. In addition, the adjoint $L_{ab}^{\dagger}$ is now calculated with respect to the new scalar product (\ref{eq:sp}) and relations (\ref{eq:com-R}) have to be taken into account, leading to
\begin{equation}
\begin{split}
  &[R_i, L_{ab}] = 0 \qquad \text{if $a,b\ne i$}, \\
  &\{R_i, L_{ab}\} = 0 \qquad \text{if $a=i$ or $b=i$}.
\end{split}
\end{equation}
\par
%
%
\section{Invariance algebra of the Dunkl-Coulomb problem in Sturm representation}

\setcounter{equation}{0}

{}From its definition (\ref{eq:K}), it is obvious that the operator $K$, which is diagonalized in Sturm representation, is a linear combination of the generators $\Gamma_0$ and $\Gamma_{d+1}$ of the so(2,1) subalgebra considered in section~3,
\begin{equation}
  K = \frac{1}{2} (1-2E) \Gamma_0 + \frac{1}{2}(1+2E) \Gamma_{d+1}.
\end{equation}
From commutation relations (\ref{eq:com-2}), it therefore follows that
\begin{equation}
  [J_{ij}, K] = [B_i, K] = 0,
\end{equation}
where $B_i$ is defined by
\begin{equation}
  B_i = \frac{1}{2}[(1-2E) A_i + (1+2E) M_i],  \label{eq:B-A-M} 
\end{equation}
which from (\ref{eq:gen}) amounts to
\begin{align}
  B_i &= \frac{1}{2} \biggl[- x_i \mathbf{D}^2 + 2 D_i \biggl(\mathbf{x}\cdot\mathbf{D} + \frac{d-3}{2}
      + \sum_j \mu_j D_j\biggr) + 2E x_i \biggr] \nonumber \\
  &= \frac{1}{2} \biggl[- x_i \mathbf{D}^2 + 2 \biggl(\mathbf{x}\cdot\mathbf{D} + \frac{d-1}{2}
      + \sum_j \mu_j D_j\biggr) D_i + 2E x_i \biggr].  \label{eq:B}
\end{align}
Hence, the $\frac{1}{2}d(d+1)$ operators $J_{ij} = - J_{ji}$ and $B_i$ are integrals of motion of the Dunkl-Coulomb problem in Sturm representation.\par
%
%
Their commutation relations directly follow from equations (\ref{eq:com-3}) and (\ref{eq:com-4}) and they are given by
\begin{equation}
\begin{split}
  &[J_{ij}, J_{kl}] = {\rm i} [\delta_{i,k} J_{jl} (1 + 2\mu_k R_k) + \delta_{i,l} J_{kj} (1 + 2\mu_l R_l) \\
  &\phantom{[J_{ij}, J_{kl}]}\quad + \delta_{j,k} J_{li} (1 + 2 \mu_k R_k) + \delta_{j,l} J_{ik} 
      (1+ 2 \mu_l R_l)],\\
  &[J_{ij}, B_k] = {\rm i} (\delta_{i,k} B_j - \delta_{j,k} B_i) (1 + 2\mu_k R_k), \\
  &[B_i, B_j] = -2{\rm i} E J_{ij}.
\end{split} \label{eq:com-B}
\end{equation}
We also get a direct consequence of (\ref{eq:com-R}), namely
\begin{equation}
\begin{split}
  &[R_i, J_{jk}] = 0 \qquad \text{if $i\ne j, k$}, \\
  &\{R_i, J_{jk}\} = 0 \qquad \text{if $i=j$ or $i=k$}, \\
  &[R_i, B_j] = 0 \qquad \text{if $i\ne j$}, \\
  &\{R_i, B_j\} = 0 \qquad \text{if $i=j$}. 
\end{split}
\end{equation}
\par
%
%
In addition, the operators $J_{ij}$ and $B_i$ satisfy the relations
\begin{equation}
  J_{ij} B_k + J_{jk} B_i + J_{ki} B_j = 0 , \qquad 1 \le i < j < k \le d,  \label{eq:J-B}
\end{equation}
and
\begin{equation}
  \mathbf{B}^2 = K^2 + 2E \biggl[\mathbf{J}^2 + \biggl(\frac{d-1}{2} + \sum_i \mu_iR_i\biggr)^2
  \biggr].  \label{eq:B-2}
\end{equation}
The proof of equation (\ref{eq:J-B}) is a straightforward consequence of definitions (\ref{eq:J}) and (\ref{eq:B}). To demonstrate equation (\ref{eq:B-2}), let us first calculate the square of a component of $\mathbf{B}$, for instance $B_1$. From definition (\ref{eq:B}), as well as from commutation or anticommutation relations (\ref{eq:Dunkl-com}), one gets
\begin{align}
  B_1^2 &= \frac{1}{4} x_1^2 \mathbf{D}^4 - \biggl[x_1\biggl(\mathbf{x}\cdot\mathbf{D} + \frac{d+1}{2}
      \biggr) D_1  + \frac{1}{2} \biggl(\mathbf{x}\cdot\mathbf{D} + \frac{d-1}{2}\biggr) + E x_1^2
      \biggr] \mathbf{D}^2 \nonumber \\
  &\quad {}- x_1 D_1 \mathbf{D}^2 \biggl(- \mu_1 R_1 + \sum_{j\ne 1} \mu_j R_j\biggr) 
      - \biggl(\mathbf{x}\cdot\mathbf{D} + \frac{d-1}{2}\biggr) \mathbf{D}^2 \mu_1 R_1 \nonumber \\
  &\quad {}- \frac{1}{2} \mathbf{D}^2 \biggl(\sum_j \mu_j R_j\biggr) (1 + 2\mu_1 R_1) + \biggl(
      \mathbf{x}\cdot\mathbf{D} + \frac{d-1}{2}\biggr) \biggl(\mathbf{x}\cdot\mathbf{D} + \frac{d+1}{2}
      \biggr) D_1^2 \nonumber \\
  &\quad {}+ (2\mathbf{x}\cdot\mathbf{D} + d) D_1^2 \sum_j \mu_j R_j + D_1^2 \biggl(\sum_j \mu_j
      R_j\biggr)^2 \nonumber \\
  &\quad {}+ 2E x_1 \biggl(\mathbf{x}\cdot\mathbf{D} + \frac{d-1}{2}\biggr) D_1 + 2E x_1 D_1
      \biggl(- \mu_1 R_1 + \sum_{j\ne 1} \mu_j R_j\biggr) \nonumber \\
  &\quad {}+ E \biggl(\mathbf{x}\cdot\mathbf{D} + \frac{d-1}{2} + \sum_j \mu_j R_j\biggr)
      (1+2\mu_1 R_1) + E^2 x_1^2. 
\end{align}
A similar calculation leads to the square of the other components of $\mathbf{B}$. On summing all the results, one obtains
\begin{align}
  \mathbf{B}^2 &= \frac{1}{4} \biggl\{\mathbf{x}^2 \mathbf{D}^4 + (2\mathbf{x}\cdot\mathbf{D}
      + d-1 - 4E\mathbf{x}^2) \mathbf{D}^2 + 2 \mathbf{D}^2 \sum_i \mu_i R_i \nonumber \\
  &\quad {}+ 4E \biggl[2(\mathbf{x}\cdot\mathbf{D})^2 + (2d-3) \mathbf{x}\cdot\mathbf{D}
      + \frac{d(d-1)}{2}\biggr] + 4E (4\mathbf{x}\cdot\mathbf{D} + 2d-1) \sum_i \mu_i R_i \nonumber \\
  &\quad {}+ 8E \biggl(\sum_i \mu_i R_i\biggr)^2 + 4E^2 \mathbf{x}^2\biggr\}.
      \label{eq:B-2-bis}
\end{align}
On the other hand, the square of operator (\ref{eq:K}) is given by
\begin{align}
  K^2 &= \frac{1}{4} \biggl\{\mathbf{x}^2 \mathbf{D}^4 + 2 \biggl(\mathbf{x}\cdot\mathbf{D}
      + \frac{d-1}{2} + \sum_i \mu_i R_i + 2E \mathbf{x}^2 \biggr) \mathbf{D}^2 \nonumber \\
  &\quad {}+ 4E \biggl(\mathbf{x}\cdot\mathbf{D} + \frac{d-1}{2} + \sum_i \mu_i R_i \biggr)
      + 4E^2 \mathbf{x}^2\biggr\}.  \label{eq:K-2}
\end{align}
It only remains to combine equations (\ref{eq:J-2}), (\ref{eq:B-2-bis}), and (\ref{eq:K-2}) to complete the proof of equation (\ref{eq:B-2}).\par
%
%
The results obtained here for the invariance algebra of the $d$-dimensional Dunkl-Coulomb problem in Sturm representation generalize those previously obtained by another approach for the three-dimensional standard Coulomb problem \cite{turbiner}. It is worth observing, however, that although equations (\ref{eq:com-B}), (\ref{eq:J-B}), and (\ref{eq:B-2}) give back equations (10), (11), and (12) of Ref.~\cite{turbiner} for $d=3$ and $\mu_i=0$, definition (\ref{eq:B}) of $B_i$ would lead to $B_i = \frac{1}{2}[x_i \mathbf{p}^2 - 2p_i (\mathbf{x}\cdot\mathbf{p}) + 2E x_i]$, which differs from definition (9) given in \cite{turbiner}.\par
%
%
\section{Invariance algebra of the Dunkl-Coulomb problem in Schr\"odinger representation}

\setcounter{equation}{0}

On considering 
\begin{equation}
  H = \frac{1}{r} (K-\alpha) + E. \label{eq:H-K}
\end{equation}
instead of $K$, it is obvious that the angular momentum components $J_{ij}$ remain integrals of motion:
\begin{equation}
  [J_{ij}, H] = 0.
\end{equation}
This is not the case, however, for the components of $\mathbf{B}$, which have to be transformed. As we plan to show now, the operators $\tilde{A}_i$, defined by
\begin{equation}
  \tilde{A}_i = B_i + x_i (H-E),  \label{eq:A-B}
\end{equation}
have that property, namely
\begin{equation}
  [\tilde{A}_i, H] = 0.
\end{equation}
From (\ref{eq:H-K}) and (\ref{eq:A-B}), this amounts to showing that
\begin{equation}
  [B_i, H] + [x_i, H] (H-E) = 0.  \label{eq:B-H}
\end{equation}
Since
\begin{equation}
  [B_i, H] = \biggl[B_i, \frac{1}{r}(K-\alpha) + E\biggr] = \biggl[B_i, \frac{1}{r}\biggr] (K-\alpha) =
  \biggl[B_i, \frac{1}{r}\biggr] r (H-E)
\end{equation}
and
\begin{equation}
  [x_i, H] = D_i,  \label{eq:x-H}
\end{equation}
equation (\ref{eq:B-H}) is equivalent to
\begin{equation}
  \biggl[B_i, \frac{1}{r}\Biggr] r = - D_i.  \label{eq:B-H-bis}
\end{equation}
\par
%
%
On inserting definition (\ref{eq:B}) in the left-hand side of (\ref{eq:B-H-bis}), we get
\begin{equation}
  \biggl[B_i, \frac{1}{r}\biggr] = \biggl[- \frac{1}{2} x_i \mathbf{D}^2 + D_i \biggl(\mathbf{x}\cdot
  \mathbf{D} + \frac{d-3}{2} + \sum_j \mu_j R_j\biggr), \frac{1}{r}\biggr],
\end{equation}
where
\begin{equation}
\begin{split}
  &\biggl[D_i, \frac{1}{r}\biggr] = - \frac{x_i}{r^3}, \\
  &\biggl[\mathbf{D}^2, \frac{1}{r}\Biggr] = - \frac{2}{r^3} \biggl(\mathbf{x}\cdot\mathbf{D}
      + \frac{d-3}{2} + \sum_j \mu_j R_j\biggr), \\
  &\biggl[\mathbf{x}\cdot\mathbf{D}, \frac{1}{r}\biggr] = - \frac{1}{r}.
\end{split}
\end{equation}
Hence, equation (\ref{eq:B-H-bis}) is fulfilled, so that the components of $\mathbf{\tilde{A}}_i$, defined in (\ref{eq:A-B}) are the searched for integrals of motion.\par
%
%
On combining equation (\ref{eq:A-B}) with (\ref{eq:H}) and (\ref{eq:B}), we get an explicit expression for the deformed LRL vectoir components
\begin{align}
  \tilde{A}_i &= - x_i \mathbf{D}^2 + D_i \biggl(\mathbf{x}\cdot\mathbf{D} + \frac{d-3}{2} + \sum_j
      \mu_j R_j\biggr) - \alpha \frac{x_i}{r} \nonumber \\
  &= - x_i \mathbf{D}^2 + \biggl(\mathbf{x}\cdot\mathbf{D} + \frac{d-1}{2} + \sum_j \mu_j R_j\biggr) 
      D_i - \alpha \frac{x_i}{r},
\end{align}
where $i=1, 2, \ldots, d$. It is worth observing that for $d=3$ and $\mu_i = 0$, we get back the usual expression for the LRL vector components $\tilde{A}_i = x_i \mathbf{p}^2 - p_i (\mathbf{x}\cdot\mathbf{p}) - \alpha \frac{x_i}{r} = \frac{1}{2} [(\mathbf{p}\times\mathbf{J})_i - (\mathbf{J}\times\mathbf{p})_i] - \alpha \frac{x_i}{r}$ with $\mathbf{J} = \mathbf{x} \times \mathbf{p}$ (see, e.g., equation (3) of Ref.~\cite{turbiner}. \par
%
%
The commutation relations of the integrals of motion $J_{ij}$ and $\tilde{A}_i$ in Schr\"odinger representation among themselves can be directly derived from those of the integrals of motion $J_{ij}$ and $B_i$ in Sturm representation, given in equation (\ref{eq:com-B}), and are given by
\begin{equation}
\begin{split}
  &[J_{ij}, J_{kl}] = {\rm i} [\delta_{i,k} J_{jl} (1 + 2\mu_k R_k) + \delta_{i,l} J_{kj} (1 + 2\mu_l R_l) \\
  &\phantom{[J_{ij}, J_{kl}]}\quad + \delta_{j,k} J_{li} (1 + 2 \mu_k R_k) + \delta_{j,l} J_{ik} 
      (1+ 2 \mu_l R_l)],\\
  &[J_{ij}, \tilde{A}_k] = {\rm i} (\delta_{i,k} \tilde{A}_j - \delta_{j,k} \tilde{A}_i) (1 + 2\mu_k R_k), \\
  &[\tilde{A}_i, \tilde{A}_j] = -2{\rm i} H J_{ij}.
\end{split}. \label{eq:com-A}
\end{equation}
To prove the last relation, use is made of
\begin{equation}
  [\tilde{A}_i, \tilde{A}_j] = [B_i + x_i(H-E), B_j + x_j(H-E)],
\end{equation}
where $[B_i, B_j]$ is already known,
\begin{equation}
  [x_i(H-E), x_j(H-E)] = - {\rm i} J_{ij} (H-E),
\end{equation}
is easily shown from (\ref{eq:x-H}), and
\begin{equation}
  [B_i, x_j(H-E)] + [x_i(H-E), B_j] = - {\rm i} J_{ij} (H-E)
\end{equation}
directly results from
\begin{align}
  [B_i, x_j] &= \frac{1}{2} [(1-2E) A_i + (1+2E) M_i, M_j - A_j] \nonumber \\
  &= {\rm i} \delta_{i,j} T (1+2\mu_i R_i) - {\rm i} J_{ij},
\end{align}
where use is made of (\ref{eq:gen}), (\ref{eq:com-3}), (\ref{eq:com-4}), and (\ref{eq:B-A-M}).\par
%
%
We also get some commutation or anticommutation relations involving reflection operators,
\begin{equation}
\begin{split}
  &[R_i, J_{jk} ] = 0 \qquad \text{if $i\ne j, k$}, \\
  &\{R_i, J_{jk}\} = 0 \qquad \text{if $i=j$ or $i=k$}, \\
  &[R_i, \tilde{A}_j] = 0 \qquad \text{if $i \ne j$}, \\
  &\{R_i, \tilde{A}_j\} = 0 \qquad \text{if $i=j$},
\end{split}
\end{equation}
as well as some additional relations
\begin{equation}
  J_{ij} \tilde{A}_k + J_{jk} \tilde{A}_i + J_{ki} \tilde{A}_j = 0, \qquad 1 \le i < j < k \le d,
  \label{eq:rel-1}
\end{equation}
and 
\begin{equation}
  \mathbf{\tilde{A}}^2 = 2H \biggl[\mathbf{J}^2 + \biggl(\frac{d-1}{2} + \sum_i \mu_i R_i\biggr)^2
  \biggr] + \alpha^2.  \label{eq:rel-2}
\end{equation}
\par
%
%
The demonstration of (\ref{eq:rel-1}) is straightforward. To prove (\ref{eq:rel-2}), one may write
\begin{equation}
  \mathbf{\tilde{A}}^2 = \mathbf{B}^2 + \mathbf{x}(H-E) \cdot \mathbf{x}(H-E)  + \mathbf{B} \cdot
  \mathbf{x}(H-E) + \mathbf{x}(H-E) \cdot \mathbf{B},
\end{equation}
where $\mathbf{B}^2$ is given in (\ref{eq:B-2-bis}), while the remaining terms can be easily shown to be given by
\begin{align}
  &\mathbf{x}(H-E) \cdot \mathbf{x}(H-E)= \frac{1}{4} \mathbf{x}^2 \mathbf{D}^4 + \frac{1}{2}
      (\mathbf{x} \cdot \mathbf{D}) \mathbf{D}^2 + \biggl(\mathbf{x}^2 \mathbf{D}^2 +
      2 \mathbf{x} \cdot \mathbf{D} \nonumber \\
  &\quad + \frac{d-1}{2} + \sum_i \mu_i R_i\biggr) \frac{\alpha}{r} + \alpha^2 + E(\mathbf{x}^2
      \mathbf{D}^2 + \mathbf{x} \cdot \mathbf{D} + 2\alpha r) + E^2 \mathbf{x}^2,
\end{align}
and
\begin{align}
  &\mathbf{B} \cdot \mathbf{x}(H-E) + \mathbf{x}(H-E) \cdot \mathbf{B} \nonumber \\
  & \quad = \frac{1}{2} \mathbf{x}^2 \mathbf{D}^4 - \biggl[(\mathbf{x} \cdot \mathbf{D})^2
       + \biggl(2 \mathbf{x} \cdot \mathbf{D} + \frac{d}{2} + \sum_i \mu_i R_i\biggr) \biggl(
       \frac{d-1}{2} + \sum_j \mu_j R_j\biggr)\biggr] \mathbf{D}^2 \nonumber \\ 
  &\quad + \biggl[\mathbf{x}^2 \mathbf{D}^2 - 2 (\mathbf{x} \cdot \mathbf{D})^2 - 2
       \biggl(2\mathbf{x} \cdot \mathbf{D} + \frac{d}{2} + \sum_i \mu_i R_i\biggr)\biggl(
       \frac{d-1}{2} + \sum_j \mu_j R_j\biggr)\biggr] \frac{\alpha}{r} \nonumber \\
  &\quad + E \biggl[-2(\mathbf{x} \cdot \mathbf{D})^2 - 2\biggl(2 \mathbf{x} \cdot \mathbf{D}
       + \frac{d}{2} + \sum_i \mu_i R_i\biggr) \biggl(\frac{d-1}{2} + \sum_j \mu_j R_j\biggr)
       - 2\alpha r\biggr] \nonumber \\
  &\quad - 2 E^2 \mathbf{x}^2.
\end{align}
On summing the three terms, we get
\begin{align}
  \mathbf{\tilde{A}}^2 &= \biggl[- \mathbf{x}^2 \mathbf{D}^2 + (\mathbf{x} \cdot \mathbf{D})^2
      + \mathbf{x} \cdot \mathbf{D} \biggl(d-2 + 2 \sum_i \mu_i R_i\biggr) + \biggl(\frac{d-1}{2}
      + \sum_i \mu_i R_i\biggr)^2\biggr] \nonumber \\
  &\quad {} \times \biggl(- \mathbf{D}^2 - \frac{2\alpha}{r}\biggr) + \alpha^2,
\end{align}
which amounts to (\ref{eq:rel-2}) on using (\ref{eq:H}) and (\ref{eq:J-2}).\par
%
%
One notes again that for $d=3$ and $\mu_i=0$, the commutation relations (\ref{eq:com-A}), as well as the additional relations (\ref{eq:rel-1}) and (\ref{eq:rel-2}), reduce to known results for the three-dimensional Coulomb problem (see, e.g., equations (4), (5), and (6) in Ref.~\cite{turbiner}).\par
%
%
\section{Conclusion}

In the present work, we have shown that the rich algebraic structure of the standard three-dimensional (or, more generally, $d$-dimensional) Coulomb problem can be extended to its Dunkl counterpart.\par
%
%
To start with, we have proved that the replacement of standard derivatives by Dunkl ones transforms the so($d+1$,2) dynamical algebra of the standard $d$-dimensional Coulomb problem into a deformed algebra with similar commutation relations, except that the metric tensor becomes dependent on the reflection operators and that there are some additional commutation or anticommutation relations involving the latter.\par
%
%
We have then shown that from some of the dynamical algebra generators, it is straightforward to derive the integrals of motion of the Dunkl-Coulomb problem in Sturm representation.\par
%
%
{}From the latter, we have then built the components of a deformed LRL vector, which, with the Dunkl angular momentum, provide integrals of motion of the Dunkl-Coulomb problem in Schr\"odinger representation and insure its superintegrability.\par
%
%
Studying the detailed action of the operators introduced in the present paper on the Dunkl-Coulomb Hamiltonian eigenstates would be an interesting subject for future work.\par
%
%
\section*{Data availability statement}

No new data were created or analyzed in this study.\par
%
%
\section*{Acknowledgment}

The author was supported by the Fonds de la Recherche Scientifique - FNRS under Grant No.~4.45.10.08.\par
%
%


\newpage

\begin{thebibliography}{99}

\bibitem{wigner}
E.\ P.\ Wigner:
Do the equations of motion determine the quantum mechanical commutation relations?,
{\sl Phys.\ Rev.} {\bf 77}, 711 (1950).

\bibitem{yang}
L.\ Yang:
A note on the quantum rule of the harmonic oscillator,
{\sl Phys.\ Rev.} {\bf 84}, 788 (1951).

\bibitem{dunkl89}
C.\ F.\ Dunkl:
Differential-difference operators associated to reflection groups,
{\sl Trans.\ Am.\ Math.\ Soc.} {\bf 311}, 167 (1989).

\bibitem{dunkl14}
C.\ F.\ Dunkl and Y.\ Xu:
{\sl Orthogonal Polynomials of Several Variables (Encyclopedia of Mathematics and its Applications)} 2nd edn,
Cambridge University Press, Cambridge, 2014.

\bibitem{plyu}
M.\ S.\ Plyushchay:
Deformed Heisenberg algebra, fractional spin fields, and supersymmetry without fermions,
{\sl Ann.\ Phys., NY} {\bf 245}, 339 (1996).

\bibitem{cq21}
C.\ Quesne:
Minimal bosonization of double-graded quantum mechanics,
{\sl Mod.\ Phys.\ Lett.\ A} {\bf 36}, 2150238 (2021).

\bibitem{brink}
L.\ Brink, T.\ H.\ Hansson and M.\ A.\ Vasiliev:
Explicit solutions of the $N$-body Calogero problem,
{\sl Phys.\ Lett.\ B} {\bf 286}, 109 (1992).

\bibitem{poly}
A.\ P.\ Polychronakos:
Exchange operator formalism for integrable systems of particles,
{\sl Phys.\ Rev.\ Lett.} {\bf 69}, 703 (1992).

\bibitem{cq95}
C.\ Quesne:
Exchange operators and the extended Heisenberg algebra for the three-body Calogero-Marchioro-Wolfes problem,
{\sl Mod.\ Phys.\ Lett.\ A} {\bf 10}, 1323 (1995).

\bibitem{cq10}
C.\ Quesne:
Superintegrability of the Tremblay-Turbiner-Winternitz quantum Hamiltonians on a plane for odd $k$,
{\sl J.\ Phys.\ A} {\bf 43}, 082001 (2010).

\bibitem{genest13a}
V.\ X.\ Genest, M.\ E.\ H.\ Ismail, L.\ Vinet and A.\ Zhedanov:
The Dunkl oscillator in the plane: I.\ Superintegrability, separated wavefunctions and overlap coefficients,
{\sl J.\ Phys.\ A} {\bf 46}, 145201 (2013).

\bibitem{genest14a}
V.\ X.\ Genest, M.\ E.\ H.\ Ismail, L.\ Vinet and A.\ Zhedanov:
The Dunkl oscillator in the plane II: Representations of the symmetry algebra, 
{\sl Commun.\ Math.\ Phys.} {\bf 329}, 999 (2014).

\bibitem{genest13b}
V.\ X.\ Genest, L.\ Vinet and A.\ Zhedanov:
The singular and the 2:1 anisotropic Dunkl oscillators in the plane,
{\sl J.\ Phys.\ A} {\bf 46}, 325201 (2013).

\bibitem{genest14b}
V.\ X.\ Genest, L.\ Vinet and A.\ Zhedanov:
The Dunkl oscillator in three dimensions,
{\sl J.\ Phys.\ Conf.\ Ser.} {\bf 512}, 012010 (2014).

\bibitem{genest15}
V.\ X.\ Genest, A.\ Lapointe and L.\ Vinet:
The Dunkl-Coulomb problem in the plane,
{\sl Phys.\ Lett.\ A} {\bf 379}, 923 (2015).

\bibitem{gha19}
S.\ Ghazouani, I.\ Sboui, M.\ A.\ Amdouni and B.\ E.\ H.\ Rhouma:
The Dunkl-Coulomb problem in three-dimensions: energy spectrum, wave functions and $h$-spherical harmonics,
{\sl J.\ Phys.\ A} {\bf 52}, 225202 (2019).

\bibitem{gha20}
S.\ Ghazouani and I.\ Sboui:
Superintegrability of the Dunkl-Coulomb problem in three-dimensions,
{\sl J.\ Phys.\ A} {\bf 53}, 035202 (2020).

\bibitem{cq23a}
C.\ Quesne:
Rationally-extended Dunkl oscillator on the line,
{\sl J.\ Phys.\ A} {\bf 56}, 265203 (2023).

\bibitem{cq23b}
C.\ Quesne:
Rational extensions of the Dunkl oscillator in the plane and exceptional orthogonal polynomials,
{\sl Mod.\ Phys.\ Lett.\ A} {\bf 38}, 2350108 (2023).

\bibitem{cq24}
C.\ Quesne:
Quasi-exactly solvable potentials in Wigner-Dunkl quantum mechanics,
{\sl Eur.\ Phys.\ Lett.} {\bf 145}, 62001 (2024).

\bibitem{wybourne}
B.\ G.\ Wybourne:
{\sl Classical Groups for Physicists}, Wiley, New York, 1974.

\bibitem{barut}
A.\ O.\ Barut and G.\ L.\ Bornzin:
SO(4,2)-formulation of the symmetry breaking in relativistic Kepler problems with or without magnetic charges,
{\sl J.\ Math.\ Phys.} {\bf 12}, 841 (1971).

\bibitem{dhoker}
E.\ D'Hoker and L.\ Vinet:
Spectrum (super-)symmetries of particles in a Coulomb potential,
{\sl Nucl.\ Phys.\ B} {\bf 260}, 79 (1985).

\bibitem{pauli}
W.\ Pauli:
\"Uber das Wasserstoffspektrum von Standpunkt der neuen Quantummechanik,
{\sl Z.\ Phys.} {\bf 36}, 336 (1926).

\bibitem{turbiner}
A.\ V.\ Turbiner and A.\ M.\ Escobar-Ruiz:
Two-body Coulomb problem and hidden $g^{(2)}$ algebra: superintegrability and cubic polynomial algebra,
{\sl J.\ Phys.\ Conf.\ Ser.} {\bf 2667}, 012075 (2023).

\end{thebibliography}
\end{document}